\documentclass[a4paper,10pt,oneside,final,onecolumn]{article}

\usepackage[utf8]{inputenc}
\usepackage{hyperref}
\usepackage[margin=3cm]{geometry}

\title{Contemporary Internet of Things platforms}
\author{Julien Mineraud, Oleksiy Mazhelis, Xiang Su and Sasu Tarkoma}
\date{28/01/2015}

\newcounter{IoTPlatform}
\newcommand{\iotplatform}[2]{\vspace{1em}
\par\noindent\refstepcounter{IoTPlatform}Platform
\arabic{IoTPlatform}: \textbf{#1} (\url{#2})\par}
\newcommand{\iotplatformbreak}[2]{\vspace{1em}
\par\noindent\refstepcounter{IoTPlatform}Platform
\arabic{IoTPlatform}: \textbf{#1}\\
(\url{#2})\par}
\newcommand{\citeiot}[1]{[Platform~\ref{#1}]}

\begin{document}

\maketitle

\begin{abstract}
 This document regroups a representative, but non-exhaustive, list of contemporary
IoT platforms. The platforms are ordered alphabetically.
 The aim of this document is to provide the a quick review of current IoT
platforms, as well as relevant information.
\end{abstract}

\iotplatform{AirVantage\textsuperscript{\texttrademark}}{https://airvantage.net/}
\label{iot:airvantage} 
AirVantage\textsuperscript{\texttrademark} is a proprietary cloud-based M2M
dedicated platform that provides end-to-end solutions to connect wireless
enabled devices to their platform. 
From an user viewpoint, the platform proposes interactive dashboards for device
management, and big data storage. The platform uses open source M2M dedicated
development tools such as the framework
\emph{m2m.eclipse.org}\footnote{\url{http://m2m.eclipse.org}}.

\iotplatform{Arkessa}{http://www.arkessa.com/}\label{iot:arkessa}
Arkessa is a proprietary cloud-based M2M management architecture and IoT
platform. It includes the MOSAIC platform that enables devices to be easily
connected to many applications. 
Privacy with third-party applications is done in similar way than Facebook or
Linkedin.  Ownership of the data remains to the end-user. Arkessa provide a
ecosystem of devices and applications providing
high flexibility to the end-user.

\iotplatform{Axeda\textsuperscript{\textregistered}}{http://www.axeda.com/}
\label{iot:axeda} 
Axeda is a proprietary cloud-based platform to enable machine-to-machine (M2M)
communication of businesses. Axeda requires the use of a proprietary messaging
protocol: 
Axeda Wireless Protocol (AWP). The platform provide SOAP and RESTful web
services that facilitate the development of applications. The platform also
provides a Groovy scripting engine to further support
the application development. A variety of tools facilitates the mashup of
assets' data and ease the visualization of this data with web-based applications
(widgets) within a dashboard.

\iotplatform{Carriots\textsuperscript{\textregistered}}
{https://www.carriots.com/}\label{iot:carriots}
Carriots\textsuperscript{\textregistered} is a proprietary cloud based platform
(PaaS). REST Api and Groovy SDK are available for web application development.
Data format supported are JSON, XML. 
The data is stored on the platform and access keys are required to access it.

\iotplatformbreak{DeviceCloud}{http://www.etherios.com/products/devicecloud/}\label{
iot:devicecloud}
DeviceCloud is a proprietary and cloud-based device management platform (PaaS).
The platform provides access the devices connected to the platform via a REST
API.

\iotplatform{Devicehub.net}{http://www.devicehub.net/}\label{iot:devicehub}
Devicehub.net is a proprietary cloud-based platform which do no provide a true
REST API (using GET method to PUT data). Currently, the documentation of the
platform is too limited to provide more information.

\iotplatformbreak{Ericsson IoT-Framework}
{https://github.com/EricssonResearch/iot-framework-engine}\label{iot:ericsson}
is a PaaS that accumulates sensor data from IP networks and focuses on the
analytics and the mashing up of the data. The PaaS includes a REST API, data
storage functionalities and OpenId access control for the data. The strength of
this platform is the pub/sub mechanism, and querying of data streams, both 
from local and external data sources) to perform analytical tasks. 
The platform also include a WebUI for interaction with users.

\iotplatform{EveryAware}{http://www.everyaware.eu/}\label{iot:everyaware}
The EveryAware platform~\cite{Becker2013} provides an extendable data concept
that could be use to enhance the possibilities of sharing and fusioning data
feeds. The platform is running on a centralized server. 
This platform was the one providing the finer-granularity of
data visibility with four different levels (details, statistics, anonymous,
none). A REST API has been integrated to access the data (extendable data
models). Typical data types are JSON, XML or PNG.

\iotplatformbreak{EveryWare Device
Cloud\textsuperscript\texttrademark}
{http://www.eurotech.com/en/products/software+services/everyware+device+cloud}
\label{iot:everyware}
EveryWare Device Cloud\textsuperscript\texttrademark is a proprietary
cloud-based platform (PaaS) using a pay-as-you-go business model. 
A RESTful API supporting JSON and XML data formats, is integrated for
communication with the devices.
The sensors required to be connected to Eurotech hardware via MQTT communication
protocol to get access to the cloud. A variety of applications and tools is
available within the platform to provide full end-to-end solution.

\iotplatform{EvryThng}{http://www.evrythng.com/}\label{iot:evrythng}
EvryThng is a proprietary centralized platform  (SaaS) that provides a
persistent presence on the Web of identifiable objects (RFID, NFC, connected
objects, etc.). It allows via RESTful API
to store and retrieve metadata as well as real-time data for these objects. The
API allows fine-access grained control to easy sharing of products informations.
No search tools are available to find data feeds. Billing is done on-demand.

\iotplatform{Exosite}{http://exosite.com/}\label{iot:exosite}
Proprietary cloud-based solution (PaaS) enabling vertical markets (from devices
to IoT solution). HTTP, JSON and UDP. Libraries for binding of the REST API with
the Exosite platform are open-source, available 
under the BSD licence. 

\iotplatform{Fosstrack}{https://code.google.com/p/fosstrak/}\label{iot:fosstrack
}
Fosstrack is an closed-source SaaS platform to handle RFID devices. Electronic
Product Code (EPC) cloud have been developed on top of the Fosstrack for fast
deployments of RFID systems. Fosstrack shows that the fragmentation of the IoT
landscape is high. However, the users stores RFID data on their own database
accessed via a Tomcat server.

\iotplatform{GroveStreams}{https://grovestreams.com/}\label{iot:grovestreams}
GroveStreams proprietary cloud based solution for analytics of data from
multiple sources. It uses a REST API and JSON data format. 
GroveStreams is an open platform, in the cloud, that any organization, user or
device can take advantage of. 
GroveStreams is free for small users. Large users will only be billed for what
they use.

\iotplatformbreak{Hub-of-All-Things}
{http://hubofallthings.com/}\label{iot:hat}
The Hub-of-All-Things (H.A.T.) is multi-disciplinary project
involving numerous researchers across six universities in the United
Kingdom. The project is still in its infancy as it started only recently, in
June 2013.
The project has as a primary objective the creation of multi-sided market
platform to create new economic and business opportunities using IoT
data generated by a ``smart home''. An important feature of the H.A.T. is that
the data belongs to the individual. The H.A.T. is performed by
the home owner and identifies context information to bring potential economic
and business models.

The H.A.T. project has similarities with the IoT hub and  architecture. The two
projects try to break the verticality of
the current IoT solutions (vertical silos using proprietary technologies) in
order to bring new innovative applications,
as well as economic and business models (horizontal IoT solutions). The two
architectures enable the end-users to get control of their data, and
thus maintaining their expectations about privacy and other issues. In
particular, the H.A.T architecture defines different sorts of ``stores'', such
as a store for physical devices and two kinds of app stores (in-store and
out-store). The ``in-apps'' (owned by either residents, landlords or building 
managers) have their content enriched by local data available on the private
H.A.T owned by the home owner to become ``out-apps'' that may be used 
by external platforms. Similar to the architecture presented in this paper, the
H.A.T. architecture provide tools (e.g. design, API)
to facilitate the emergence of a new kind of market, that currently does not
exists, and relies on the power of the IoT.

\iotplatform{IFTTT}{https://ifttt.com/}\label{iot:ifttt} (``if this then that'')
is a SaaS offering, allowing a rapid composition of services called ``recipes'' 
by applying simple if-then rules to external service building blocks, such as
emails, Facebook events, or Belkin's WeMo switch, that either play the role of a
trigger 
(if) or an action (then). Though the service is free to use, the APIs to the
service are not open at the time of writing. The recipes can be personal or
shared at the 
discrepancy of the user; otherwise, the service building blocks rather than
IFTTT deal with the user generated data. 

\iotplatform{LinkSmart\textsuperscript{\texttrademark}}
{http://www.hydramiddleware.eu/news.php}\label{iot:linksmart}
The LinkSmart\textsuperscript{\texttrademark} middleware platform, formerly
Hydra, is an open-source platform licensed under the LGPLv3. LGPLv3 is a
non-viral version of the GPLv3.
The platform enable the creation of a network for embedded systems, using
semantics to discover the devices connected to the network. The middleware is
based on a service-oriented
architecture. The platform provides a SDK for application development and a DDK
for device development.

\iotplatform{MyRobots}{http://www.myrobots.com/}\label{iot:myrobots}
MyRobots is a dedicated cloud-based (close) platform to connect robots to the
IoT. Data format supported are JSON, XML, CSV and the web services are buildable
using REST api. 
By default, the privacy of robots is set to public, but can be changed to
private. The platform enables robots to be controlled over the Internet. The
platform also includes an AppStore.

\iotplatform{Niagara$^{AX}$}{http://www.niagaraax.com/}\label{iot:niagara}
Niagara\textsuperscript{AX}~\cite{Samad2007} is a close proprietary M2M
dedicated software development framework that is fully distributed. 
It interconnect heterogeneous devices. However, details are missing about the
nature of the Open API.

\iotplatform{Nimbits}{http://www.nimbits.com/}\label{iot:nimbits}
Similarly to \emph{SensiNode}~\citeiot{iot:sensinode}, the Nimbits server has
been made cloud architecture compatible, hence it scales from a single private
server to a cloud architecture. 
Nimbits includes three level of private for the data: (i) private, (ii)
protected (read-only is public) and (iii) public. Control over the data and its
ownership is to the user. 
The data is transmitted via XMPP messaging protocol. Web services access the
data with HTML POST request and JSON data format. 
The platform is open source licensed under the Apache License, Version 2.0. This
license ease the integration
with GPLv3 as long as the resulting software is licensed under GPLv3.

\iotplatform{NinjaBlock}{http://ninjablocks.com/}\label{iot:ninjablock}
NinjaBlock provides open-source hardware and open-source software to facilitate
the development of sensors. However, the Ninja platform is proprietary and
cloud-based.
A RESTful API is disponible to connect NinjaBlock hardware to the cloud.
NinjaBlock is open-hardware and serves as a gateway between the sensors and the
Ninja platform. 
JSON data format is used by the platform and access is granted via the OAuth2
authentification protocol.

\iotplatform{OpenIoT}{http://openiot.eu/}\label{iot:openiot}
OpenIoT platform is an open-source platform, fully decentralized, that provides
connectivity with constrained devices such as sensors. The platform provides a
billing mechanism
for the use of services.

\iotplatform{OpenRemote}{http://www.openremote.org}\label{iot:openremote}
OpenRemote is a centralized open-source platform, licensed under the Affero GNU
Public License where the copyleft of each license is relaxed to allow
distribution of combinations with GPL for the latest versions of the licenses).
The platform supports home and domotic automation spaces using a top-down
approach.

\iotplatform{Open.Sen.se}{http://open.sen.se/}\label{iot:opensense}
Open.Sen.se is currently in a closed beta version  (PaaS/SaaS). A tool called
\emph{Funnel} can be used to aggregate data, but only on data feeds that are
within our dashboard. 
It is possible to get the data from different source and mash it up. The
platform uses the JSON data format and REST API for web services development.
Device connected to the service are usually ethernet enabled. The privacy of
data visualization is either public or private, data is always private (needs
private keys at all times to use the API). No billing is yet available.

\iotplatform{realTime.io}{https://www.realtime.io/}\label{iot:realtimeio}
IoBridge realTime.io provides a proprietary cloud based platform (PaaS) to
connect devices to the Internet and build applications upon the data. As
realTime.io uses a proprietary transport
protocol for data, \emph{ioDP}, the physical devices need to be connected to the
realTime.io cloud service via a proprietary gateway. Once these gateways are
connected to the service,
public API (requiring realTime.io keys) enables the connection to the device to
pull or push data to the devices. The data format supported is JSON. 
No information was available on the ownership of the data. Not possible to
access a public version of the data streams as it is with
\emph{ThingSpeak}~\citeiot{iot:thingspeak}.

\iotplatform{SensiNode}{http://www.sensinode.com/}\label{iot:sensinode}
SensiNode/ARM\textsuperscript{\textregistered} provides the NanoService
platform, that is proprietary, to connect 6LoWPAN enabled devices to the IoT. 
The NanoService platform can however be run on either a private server, a
private cloud or a public cloud.
It uses CoAP and RESTful API for creating M2M networks of highly constrained
devices. Connection to unconstrained networks (normal Internet) is made 
through a NanoRouter gateway. The platform includes the Constrained RESTful
Environments (CoRE), the equivalent of REST API for constrained devices
(lightweight).

\iotplatform{SensorCloud\textsuperscript{\texttrademark}}
{http://www.sensorcloud.com/}\label{iot:sensorcloud}
SensorCloud\textsuperscript{\texttrademark} is a proprietary cloud-based sensor
data storage and visualization platform (PaaS).  
It provides a fully REST compliant API and the CSV and XDR data formats are
supported. It also provides tools for visualization and data mashup
(MathEngine). 
Data owners can also augment their audience by sending invitations to domain
experts to view their data set, assist with analysis, and develop advanced,
custom-tailored data processing applications.

\iotplatform{SkySpark}{http://skyfoundry.com/skyspark/}\label{iot:skyspark}
SkySpark is a proprietary software that can be locally installed on a private
server or on a cloud and enable analytic tools for big data processing. 
The software does not require the connection of devices to the cloud. 
The software includes a REST API for connection with third party applications
and web services. 
The SkySpark software does not include direct management of connected devices.

\iotplatform{Swarm}{http://buglabs.net/products/swarm}\label{iot:bugswarm}
Bug's Swarm cloud-based platform (PaaS) is not open-source but provides an
open-source client and some tools (unknown licence). 
It creates swarm of resources to consume data, produce data or both amongst
actors connected to the swarm.
There is limited information on how the swarm data is stored, and who had its
ownership.
A RESTful API and JSON data format are usable to communicate
with the devices. The platforms also provide GUI tools, such an interactive
dashboard with data visualization capabilities.

\iotplatform{TempoDB}{https://github.com/tempodb}\label{iot:tempodb}
TempoDB is a proprietary, cloud-based PaaS that enables the users to upload
their data on the cloud via a REST API.
The service enables to store, retrieve, and query the data, while ensuring data
security, multiple back-ups and providing visualization tools, etc. 
This service offers billing offers depending on the user need. Theses services
are used by the NinjaBlocks.

\iotplatform{TerraSwarm}{http://www.terraswarm.org/}\label{iot:terraswarm}
Similarly to the H.A.T. project~\citeiot{iot:hat}, the TerraSwarm
project~\cite{Lee2012} is a multi-disciplinary project in its infancy, which
started at the beginning
of the year 2013. The TerraSwarm project has initiated by the TerraSwarm
Research Center which will be headquartered in the University of California
Berkeley. Unlike, the H.A.T project, the vision of TerraSwarm is not limited to
the home space, but extends to ``smart cities''. The project envision
the development of a new kind of operating system, the SwarmOS, to natively
support the heterogeneous nature of the devices and solutions existing
in the IoT and enable the infrastructure with the ability to aggregate
information from a variety of data sources. The architecture relies heavily
on the power of cloud computing. The operating system will be also open-source
to improve its reliability and efficiency, while maximizing the potential
of innovative development of ``swarm-apps'' build upon the system.

\iotplatform{The thing
system}{http://thethingsystem.com/}\label{iot:thethingsystem}
The thing system is a software using \emph{node.js} that enables discovery of
smart things in the home environment. 
The project is open-source and licensed under the M.I.T license. The software
does not provide
storage functionalities and must be coupled with a PaaS to enable storage
outside the home area. 
The software intends only to provide access remotely to smart devices of smart
homes.

\iotplatform{Thing
Broker}{http://www.magic.ubc.ca/wiki/pmwiki.php/ThingBroker/ThingBroker}\label{
iot:thingbroker}
The Thing Broker~\cite{PerezdeAlmeida2013} is extending the Magic Broker 2
(MB2)~\cite{Blackstock2010} platform. It is also available as open-source. 
The centralized platform provides a Twitter-based abstraction model for
\emph{Things} and \emph{Events}, that could be used to create local ecosystems
such as smart homes. A REST API is provided by the platform to access the data 
and devices.

\iotplatform{ThingSpeak}{https://www.thingspeak.com/}\label{iot:thingspeak}
ThingSpeak is decentralized, open-source and copyrighted by ioBridge under the
license GPLv3. Commercial software or hardware using ThingSpeak requires a
commercial agreement with IoBridge Inc. 
ThingSpeak provides a server that may be used to store and retrieve IoT data. 
It allows opening of the channels (data flows, support the JSON, XML, CSV
data formats) to the public but do not provide extensive configuration of the
data flows. The platform also provides visualization tools 
and enables the creation of widgets in Javascript/HTML/CSS to visualize the data
in a more personified fashion.

\iotplatform{ThingSquare}{http://thingsquare.com/}\label{iot:thingsquare}
ThingSquare is a proprietary cloud-based platform specialized on connecting
constrained devices. It require a gateway, but its firmware is open source. 
The gateway creates a wireless mesh networks of sensors and connect it to the
Internet. The devices can access the Internet, but the devices are
invisible from outside the mesh. The platform also includes a protocol for
constrained devices. 

\iotplatform{ThingWorx}{http://www.thingworx.com/}\label{iot:thingworx}
ThingWorx is a proprietary cloud-based M2M dedicated platform (PaaS). It
provides a variety of tools and services to support end-to-end solutions. The
devices and data are accessible
via a REST API. The offer is similar to Axeda's~\citeiot{iot:axeda}.

\iotplatform{Sense Tecnic
WoTkit}{http://sensetecnic.com/}\label{iot:webofthingshub}
The WoTkit~\cite{Blackstock2012} is a proprietary cloud-based platform that
offers an interesting search tool for public sensor. Public sensors do not
require an account to be used.
The platform uses probably the open source platform Magic Broker
2~\cite{Blackstock2010} for internal operations.

\iotplatform{Xively}{https://xively.com/}\label{iot:xively}
Xively (formerly Pachube) is a proprietary cloud-based platform (PaaS).
Ownership of the data remains to the user, but the data is stored on the Xively
server. 
Xively provides open-source APIs (in various programming languages) mostly with
the BSD 3-clause licence which is very permissive licence. However, 
these libraries are rather small and do not provide great help in manipulating
the Xively API. Xively supports JSON, XML and CSV data format. Xively provides
an extensive
RESTful API including a search tool in order to retrieve feeds (flow of data)
depending on selected characteristics (location radius, name, type of data
stored, etc.)

\bibliographystyle{plain}
\bibliography{techreport_contemporary_iot_platforms}

\begin{thebibliography}{1}

\bibitem{Becker2013}
Martin Becker, Juergen Mueller, Andreas Hotho, and Gerd Stumme.
\newblock A generic platform for ubiquitous and subjective data.
\newblock In {\em Proceedings of the 2013 ACM conference on Pervasive and
  ubiquitous computing adjunct publication}, UbiComp '13 Adjunct, pages
  1175--1182, New York, NY, USA, 2013. ACM.

\bibitem{Blackstock2010}
M.~Blackstock, N.~Kaviani, R.~Lea, and A.~Friday.
\newblock Magic broker 2: An open and extensible platform for the {Internet of
  Things}.
\newblock In {\em Internet of Things (IOT)}, pages 1--8, 2010.

\bibitem{Blackstock2012}
Michael Blackstock and Rodger Lea.
\newblock Wotkit: a lightweight toolkit for the web of things.
\newblock In {\em Proceedings of the Third International Workshop on the Web of
  Things}, WOT '12, pages 3:1--3:6, New York, NY, USA, 2012. ACM.

\bibitem{Lee2012}
Edward~A. Lee, John~D. Kubiatowicz, Jan Rabaey, Alberto
  Sangiovanni-Vincentelli, Sanjit~A. Seshia, John Wawrzynek, David Blaauw,
  Prabal Dutta, Kevin Fu, Carlos Guestrin, Roozbeh Jafari, Douglas~L. Jones,
  Vijay Kumar, Richard Murray, George Pappas, Anthony Rowe, Carl Sechen,
  Tajana~Simunic Rosing, and Ben Taskar.
\newblock The terraswarm research center {(TSRC)} (a white paper).
\newblock Technical Report UCB/EECS-2012-207, EECS Department, University of
  California, Berkeley, November 2012.

\bibitem{PerezdeAlmeida2013}
Ricardo~Aparecido Perez~de Almeida, Michael Blackstock, Rodger Lea, Roberto
  Calderon, Antonio~Francisco do~Prado, and Helio~Crestana Guardia.
\newblock Thing broker: a twitter for things.
\newblock In {\em Proceedings of the 2013 ACM conference on Pervasive and
  ubiquitous computing adjunct publication}, UbiComp '13 Adjunct, pages
  1545--1554, New York, NY, USA, 2013. ACM.

\bibitem{Samad2007}
T.~Samad and B.~Frank.
\newblock Leveraging the web: A universal framework for building automation.
\newblock In {\em American Control Conference, 2007. ACC '07}, pages
  4382--4387, 2007.

\end{thebibliography}
 
\end{document}